%% file: main.tex
\documentclass[11pt]{article}
\usepackage{arxiv}

\newcommand{\R}{\mathbb{R}}
\newcommand{\E}{\mathbb{E}}
\newcommand{\N}{\mathcal{N}}

\newcommand{\TV}{\operatorname{TV}}
\newcommand{\KL}{\operatorname{KL}}

\title{Structure-dependent failure modes of neural priors in acoustic full-waveform inversion}
\author[1]{Ziye Yu\thanks{Corresponding author: \href{mailto:yuziye@cea-igp.ac.cn}{yuziye@cea-igp.ac.cn}. ORCID: \href{https://orcid.org/0000-0002-1720-3811}{0000-0002-1720-3811}.}}
\author[2,3]{Xin Liu}
\author[4]{Yuqi Cai}
\affil[1]{Institute of Geophysics, China Earthquake Administration, Beijing 100081, China}
\affil[2]{Laboratory of Seismology and Physics of Earth's Interior, School of Earth and Space Sciences, University of Science and Technology of China, Hefei 230026, China}
\affil[3]{Institute of Advanced Technology, University of Science and Technology of China, Hefei 230088, China}
\affil[4]{University of Chinese Academy of Sciences, Beijing 100029, China}
\date{}
\hypersetup{pdftitle={Structure-dependent failure modes of neural priors in acoustic full-waveform inversion},pdfauthor={Ziye Yu, Xin Liu, Yuqi Cai}}

\begin{document}

\maketitle
\begin{abstract}
Frequency continuation does not improve every neural representation, and better velocity recovery need not imply better prediction of unseen waveforms. We compare a velocity grid with total-variation regularization, an adapted sinusoidal coordinate network (SIREN), and a frozen image generator with a full-grid residual in two-dimensional acoustic full-waveform inversion. All methods use the same frozen SWEEP discretization, acquisition, observations, background, and budget of 1,500 shot--model gradient evaluations. Three exposed synthetic targets represent inclined layering, curved layering, and a high-contrast curved interface drawn from a fault-labelled family. With 2-to-10 Hz continuation, the generative-residual model attains velocity RMSEs of 218.7, 108.0, and 341.9 m/s, compared with 232.5, 100.3, and 412.9 m/s for Grid+TV. Thus the latter wins on curved layering, while the former improves model RMSE by 5.9\% and 17.2\% on the other targets. Grid+TV nevertheless predicts held-out 10 Hz shots better on all three. SIREN is worse than the common background in every cell; continuation worsens two targets but improves one. A five-weight TV sensitivity study preserves the CurveFault model-error ranking. An equal-budget variational extension fails both recovery and interval coverage. These results establish conditional rankings, not universal superiority or an isolated spectral-bias mechanism. Full-grid residuals also prevent attributing gains uniquely to the generator. We advocate joint reporting of model recovery, unseen-shot prediction, optimization cost, and negative outcomes when auditing neural FWI.
\end{abstract}
\noindent\textbf{Keywords:} full-waveform inversion; neural prior; benchmark; spectral bias; structure-dependent
\input{sections/introduction}
\input{sections/methods}
\input{sections/results}
\input{sections/discussion}
\input{sections/conclusions}
\section*{Data and code availability}
SWEEP, OpenFWI, and the original Marmousi2 model are publicly available from their cited sources. The experiment harness, frozen configurations, input checksums, generator checkpoint, saved reconstructions, and figure-generation scripts are publicly available in the \path{acoustic-fwi-structural-benchmark} directory at \url{https://huggingface.co/cangyeone/relational-geophysics/tree/main/acoustic-fwi-structural-benchmark}. The release contains the synthetic inputs, 18 primary reconstructions, TV-sensitivity and VI-8 outputs, pinned SWEEP implementation, environment requirements, and per-file integrity checks. Reproduction instructions and validation scope are documented in the project README. The release begins at the frozen generator checkpoint; original image-pretraining archives remain with their cited providers.

\section*{Acknowledgments and use of AI tools}
We acknowledge the authors and maintainers of SWEEP, Deepwave, OpenFWI, Marmousi2, SIREN, and the public image datasets. OpenAI Codex assisted with software development, experiment audits, figure preparation, and manuscript drafting. Scientific responsibility, interpretation, and final verification remain with the authors.
\clearpage

\FloatBarrier
\bibliographystyle{unsrtnat}
\bibliography{references}
\clearpage
\appendix
\input{sections/appendix}

\end{document}

%% file: sections/introduction.tex
\section{Introduction}
Full-waveform inversion (FWI) estimates subsurface properties by fitting recorded waveforms with a wave-equation model. Its resolving power is accompanied by sensitivity to initialization, acquisition aperture, frequency content, and the nonconvexity of waveform matching \citep{virieux2009,bunks1995}. A representation of the velocity field is therefore more than an implementation choice. It determines which model changes are convenient to express, how the data gradient is transformed into an update, and which structures a regularizer discourages.

Three approaches make different commitments. Direct grid inversion preserves local degrees of freedom and uses explicit penalties, such as total variation (TV), to control roughness. Implicit neural representations instead map spatial coordinates to physical properties; sinusoidal networks can represent detailed signals and their derivatives \citep{sitzmann2020}, and have motivated neural formulations of FWI \citep{sun2022}. A pretrained generator supplies a different constraint: data are explained through coordinates learned before the inverse problem is encountered. Generative priors have been studied in inverse problems \citep{bora2017}, while the deep image prior demonstrates that architecture itself can regularize reconstruction even without pretraining \citep{ulyanov2018}. These mechanisms should not be conflated.

Comparisons become difficult when a change of representation also changes the propagator, acquisition, initial model, source bandwidth, or computational expenditure. Differentiable wave packages, including SWEEP \citep{wang2026} and Deepwave \citep{richardson2026}, make it possible to compare representations within a common computational graph. Here SWEEP is the shared forward engine for \emph{all} methods; ``Grid+TV'' denotes a particular grid inversion implemented on that engine, rather than the complete capability of the SWEEP package. OpenFWI provides diverse synthetic structures for reproducible research \citep{deng2022}; borrowing its model files does not, by itself, constitute an evaluation on its official held-out split.

We present a controlled diagnostic comparison on three exposed examples, using a frozen operator and a strict accounting of shot--model gradients. The design compares direct 10 Hz inversion with 2-to-10 Hz continuation, tests the sensitivity of the grid baseline to TV strength, and distinguishes velocity-field accuracy from prediction on unused source positions. We also report an equal-budget variational extension because a successful point estimate is insufficient evidence for a useful posterior approximation.

Our contribution is an auditable set of conditional outcomes. The generative-residual parameterization has the lowest final model RMSE on two examples, Grid+TV on the third, and Grid+TV has the lowest held-out waveform error under continuation on all three. The adapted SIREN fails to improve on the background, but its response to continuation is not uniform. These observations motivate a structure-aware evaluation programme; one example per family does not yet establish a general family-level law. We separate these measurements from proposed mechanisms and from the stronger attribution experiments still needed.

%% file: sections/methods.tex
\section{Methods}
\subsection{Shared acoustic problem and discrete forward engine}
We estimate a scalar compressional velocity field $m(\mathbf{x})=V_p(\mathbf{x})$. The constant-density acoustic model has the continuum form
\begin{equation}
\frac{1}{m(\mathbf{x})^2}\frac{\partial^2 u_s}{\partial t^2}-\nabla^2u_s=f_s,
\qquad d_{s,f}=P_su_s(m;f)+\epsilon_{s,f},
\label{eq:wave}
\end{equation}
where $s$ denotes source position, $f$ denotes source-frequency experiment, and $P_s$ samples receivers. All synthesis and inversion use the same discrete SWEEP \texttt{Acoustic} equation and \texttt{PropTorch} operator \citep{wang2026}. No elastic $V_s$ or density inversion is included.

The numerical problem is fixed across targets and methods (Table~\ref{tab:protocol}). The archived installation records \texttt{sweepx} 0.2.1, \texttt{sweep-solver} 0.2.0, PyTorch 2.14.0, and NumPy 2.5.3. The executable source files, rather than package names alone, are identified by checksums. We use the CPU eager path with compiled steps, a fourth-order spatial stencil, \texttt{cpmlr} absorption, and no checkpointing. A scoped correction replaces the backward operation for replicated boundary padding with its exact vector--Jacobian product: contributions from replicated halo cells are summed onto the boundary rather than discarded. The forward padding, wave equation, stencil, and PML remain unchanged. The correction is common to every method and was checked by finite differences on a development case. Thus ``frozen native SWEEP'' means the frozen native forward implementation with this disclosed derivative correction, not an unmodified upstream backward implementation.

\begin{table}[tb]
\centering\small
\caption{Frozen physical and computational protocol. Node extent, rather than cell count times spacing, defines the stated domain. Source coordinates are $(x,z)$ grid indices; receiver indices are rounded onto the same grid.}
\label{tab:protocol}
\begin{tabularx}{\linewidth}{@{}lX@{}}
\toprule
Quantity & Value \\
\midrule
Domain and grid & $690\times690$ m; $101$ depth nodes $\times161$ horizontal nodes \\
Spacing and sampling & $\Delta z=6.9$ m, $\Delta x=4.3125$ m; $\Delta t=0.0004$ s; 4,500 samples (1.8 s) \\
Absorption & 20 PML cells; \texttt{cpmlr}; reference frequency 10 Hz; maximum velocity 5,500 m/s; no free surface \\
Sources & Five: horizontal indices $[16,49,81,114,144]$, depth index 1 \\
Receivers & 70 at depth index 1; horizontal index $\mathrm{round}(10j/4.3125)$, $j=0,\ldots,69$ \\
Shot split & Training indices $[0,2,4]$; held-out indices $[1,3]$ (zero based) \\
Wavelets & \texttt{sweep.signal.ricker}; $f=2,10$ Hz, time delays $1.5/f$ s \\
Noise & Gaussian, SD = 3\% of clean training-shot RMS separately at each frequency \\
Seeds & Network: 20260924; observation-noise draw: 20260941 \\
Budget per cell & 1,500 shot--model gradients; 501 Adam updates for point estimates \\
Continuation & 500 gradients at 2 Hz, then 1,000 at 10 Hz; model and Adam states retained \\
\bottomrule
\end{tabularx}
\end{table}

For each target we synthesize two source experiments, one with a 2 Hz Ricker wavelet and the other with a 10 Hz Ricker wavelet. These are \emph{not} frequency-filtered versions of a single broadband recording. The Gaussian standard-normal draw is shared across methods and reused across targets with target-specific noise amplitude. Within a live comparison matrix, all methods share one solver instance and one observed-data tensor. Historical and supplemental runs reuse identical archived input bytes and settings, rather than the same in-memory object across processes. The held-out shots are excluded from optimization; their final 10 Hz residuals provide a separate predictive diagnostic.

For active frequency $f$, the common data objective is
\begin{equation}
\mathcal{L}_f(m)=\frac{1}{2N_{\rm obs}}\sum_{s\in\mathcal{S}_{\rm train}}
\left\|\frac{F_{s,f}(m)-d_{s,f}}{\sigma_f}\right\|_2^2,
\qquad N_{\rm obs}=3\times4500\times70=945{,}000.
\label{eq:loss}
\end{equation}
Shot subsets use the corresponding mean loss. Most updates include all three training shots; smaller batches at the budget boundary enforce exact accounting. Each training shot is charged 500 times over a complete run. Independent wavefields may be batched in groups of at most 15, following the archived numerical audit. No target-specific change to sampling or absorption is allowed.

\subsection{Targets and the common background}
The targets are a Marmousi2 region \citep{martin2006}, \texttt{curvevel\_a\_01}, and \texttt{curvefault\_a\_01}. The latter two are previously exposed samples from official OpenFWI \emph{training} files: local \texttt{curvevel\_a\_model1.npy}, raw index 31, and \texttt{vel2\_1\_0.npy}, raw index 17. Their local case suffixes are not official dataset split identifiers. Observations are regenerated using SWEEP; we do not use the released OpenFWI seismic traces or claim an official blind benchmark result.

The Marmousi region originally spans $x=6$--10 km and $z=0$--2.5 km. Its stored $101\times161$ velocity field is anisotropically remapped to the common $690\times690$ m domain. Velocities remain in m/s; only coordinates change. This is a rescaled structural test, not native-scale Marmousi FWI. Its velocities range from 1,500 to 4,428.65 m/s, with mean 2,235.24 m/s. The two OpenFWI fields are bilinearly interpolated from $70\times70$, with aligned corners. The selected CurveFault truth primarily contains a strong curved interface; its family label should not be read as evidence of a resolved fault offset in this individual sample.

For every target, the common background $m_0$ is obtained by Gaussian smoothing at a 100 m scale in the computational domain. For OpenFWI this is $\sigma=10$ pixels on the native grid, before interpolation. Marmousi uses the equivalent physical smoothing scale on its computational grid. This is privileged synthetic initialization because it uses the truth, although every method receives exactly the same background. The background RMSEs are 275.35, 185.70, and 440.34 m/s, respectively. Each physical parameterization is anchored so that its initial central model equals $m_0$ exactly.

\subsection{Grid inversion with total variation}
Grid+TV optimizes $q\in\R^{101\times161}$ through
\begin{equation}
 m(q)=m_0+300q,\qquad
 J_{\rm grid}=\mathcal{L}_f(m(q))+\alpha\TV_\epsilon(m(q)).
\label{eq:grid}
\end{equation}
Velocity increments are in m/s. The implemented smoothed isotropic TV is
\begin{equation}
\TV_\epsilon(m)=\frac{1}{100\times160}\sum_{i,j}
\left[\sqrt{(D_zm)_{ij}^2+(D_xm)_{ij}^2+\epsilon^2}-\epsilon\right],
\label{eq:tv}
\end{equation}
where $D_zm=(10/\Delta z)(m_{i+1,j}-m_{i,j})$, $D_xm=(10/\Delta x)(m_{i,j+1}-m_{i,j})$, and $\epsilon=1$ in m/s units. TV is an edge-preserving regularizer \citep{rudin1992}; it is not equivalent to Gaussian smoothing and is not intrinsically incompatible with discontinuities.

Adam \citep{kingma2015} uses learning rate $10^{-2}$. The main weight $\alpha=10^{-3}$ was selected from $\{10^{-4},3\times10^{-4},10^{-3}\}$ on a separate exposed development sample, \texttt{curvefault\_a\_00} (raw index 3), with 300 gradients per candidate. It was then frozen. The subsequent five-weight sensitivity study is reported as a post-exposure diagnostic, without replacing the primary baseline by its most favourable outcome.

\subsection{Adapted SIREN coordinate representation}
The coordinate network has four hidden sine layers of width 256, $\omega_0=30$, input coordinates normalized to $[-1,1]^2$, and 198,401 trainable parameters. Its layer classes are extracted from the archived official SWEEP SIREN notebook. This is an adapted comparator, not a claim to reproduce every optimization choice or best result of a published IFWI method \citep{sitzmann2020,sun2022}.

Let $N_\phi(\mathbf{x})$ be its scalar output, $\phi_0$ its background-fitted parameters, $s$ the logistic function, and $b_0=\operatorname{logit}[(m_0-1200)/4300]$. We use
\begin{equation}
 m_\phi=m_0+4300\left\{s[b_0+N_\phi-N_{\phi_0}]-s(b_0)\right\}.
\label{eq:siren}
\end{equation}
This anchors the initial model exactly and bounds velocity to 1,200--5,500 m/s up to roundoff. No TV term is added. The learning rate during waveform inversion is $10^{-4}$. We call this representation spatially unregularized, rather than unconstrained: the output transform imposes physical bounds.

\subsection{Frozen generator with full-grid residual correction}
\label{sec:ours}
The generator is the decoder of a scalar variational autoencoder \citep{kingma2014}, with a 128-dimensional latent and a native $256\times256$ output. It was pretrained on generic public images from EuroSAT, BSDS500, and the Describable Textures Dataset \citep{helber2019,arbelaez2011,cimpoi2014}. The archived run used 29,816 training image identities, 2,000 separate validation identities, and 900,000 augmented training draws; no seismic velocity model entered pretraining or checkpoint selection. Appendix~\ref{app:prior} gives architecture and training details.

Let $R$ be bilinear resizing to $101\times161$ with aligned corners. The implemented anchored model is
\begin{equation}
 m(z,q)=m_0+2000R\left[G_\theta(z)-G_\theta(z_0)\right]+300q,
 \quad z\in\R^{128},\quad q\in\R^{101\times161},
\label{eq:ours}
\end{equation}
where $\theta$ and the reference $z_0$ are fixed. The residual $r=300q$ has full-grid degrees of freedom and no spatial smoothing. It nevertheless has an explicit independent Gaussian prior. The objective called Ours MAP is
\begin{equation}
 J_{\rm ours}=\mathcal{L}_f(m(z,q))+
 \frac{\|z\|_2^2+\|q\|_2^2}{2N_{\rm obs}}.
\label{eq:map}
\end{equation}
Thus ``full-grid'' does not mean an absent prior: the residual's prior scale is 300 m/s and has no imposed spatial correlation. This parameterization has 16,389 optimized coordinates, including the residual, rather than only 128.

The latent and residual learning rates are $3\times10^{-3}$ and $10^{-2}$, respectively; each parameter block has a gradient-norm clip of 10. All Adam runs use the default moment factors $(0.9,0.999)$ and $\epsilon_{\rm Adam}=10^{-8}$. Ours terminates on a nonfinite value or any total-field velocity outside the open interval $(1200,5500)$ m/s. This physical-support guard does not clip the model or replace a failed run with a background. Grid and SIREN have no corresponding reject-style guard, although numerical failures are monitored.

Before waveform inversion, $z_0$ is fitted to the background for 500 network-only gradient steps; SIREN receives 500 analogous background-fitting steps. These have zero PDE cost and are disclosed separately from the 1,500-gradient inversion budget. Anchoring in Eqs.~\eqref{eq:siren} and \eqref{eq:ours} makes the initial physical fields identical despite imperfect network background fits. Generator pretraining is also an amortized extra cost, not part of the per-target PDE budget.

Because $q$ spans the full grid, the decomposition of $m$ into generator and residual contributions is not identifiable from waveforms alone: a generator change can be cancelled by a residual change. Priors and optimization determine the allocation. The experiment therefore compares complete parameterization--regularization systems, not the generator's isolated contribution.

\subsection{Budget, evaluation, and supplementary diagnostics}
Each primary method--schedule cell receives 1,500 actual shot--model gradient evaluations, corresponding to 501 Adam updates. Direct inversion spends all of them at 10 Hz. Continued inversion spends 500 at 2 Hz and 1,000 at 10 Hz, carrying model and optimizer states forward. Equal PDE-gradient cost is not equal wall time, parameter count, or lifetime training cost.

All headline results use the final budgeted iterate, never an intermediate model chosen by truth RMSE. We compute
\begin{equation}
 E_m=\sqrt{\frac{1}{N_m}\|m-m_{\rm true}\|_2^2},\qquad
 \Delta_{\rm bg}=100\left(\frac{E_m}{E_{m_0}}-1\right),
\label{eq:metrics}
\end{equation}
with independent float64 recomputation from saved fields. Negative $\Delta_{\rm bg}$ denotes improvement. Training and held-out 10 Hz waveform scores are residual RMS divided by the known noise SD. Values near one indicate residuals on the noise scale, but do not alone establish model correctness or posterior calibration.

Supplementary analyses comprise a CurveFault TV sweep over $10^{-4}$ to $10^{-2}$, descriptive SIREN trajectory and wavenumber diagnostics, and a CurveFault mean-field VI experiment with eight Monte Carlo samples per full update (VI-8). The last jointly optimizes Gaussian means and SDs for $z$ and $q$, charging every sampled shot--model gradient. At the same 1,500-gradient cost it receives only 63 Adam updates. Its final target is the 10 Hz likelihood, with the 2 Hz stage serving as optimization continuation. Appendix~\ref{app:vi} states its objective and coverage calculation. None of these diagnostics is represented as a new blind evaluation.

%% file: sections/results.tex
\section{Results}
\subsection{Final model-error rankings depend on the target}
Table~\ref{tab:rmse} reports every primary endpoint. With continuation, Ours MAP attains the lowest velocity RMSE on the Marmousi region and CurveFault, while Grid+TV wins on CurveVel. Relative to continued Grid+TV, Ours reduces model RMSE by 5.93\% and 17.20\% in the first two cases but increases it by 7.68\% on CurveVel. These are deterministic differences for these runs, not statistical significance estimates. All 18 cells finish their budgets with finite values; no Ours physical-support guard is triggered. Poor model recovery therefore must be distinguished from numerical divergence.

\begin{table}[tb]
\centering
\caption{Final velocity RMSE in m/s after 1,500 shot--model gradients per cell. Bold denotes the lowest model RMSE for each target. All six methods start from the same target-specific background. Values are rounded only for display.}
\label{tab:rmse}
\input{data/rmse_table}
\end{table}

\begin{landscape}
\begin{figure}[p]
\centering
\includegraphics[width=.98\linewidth]{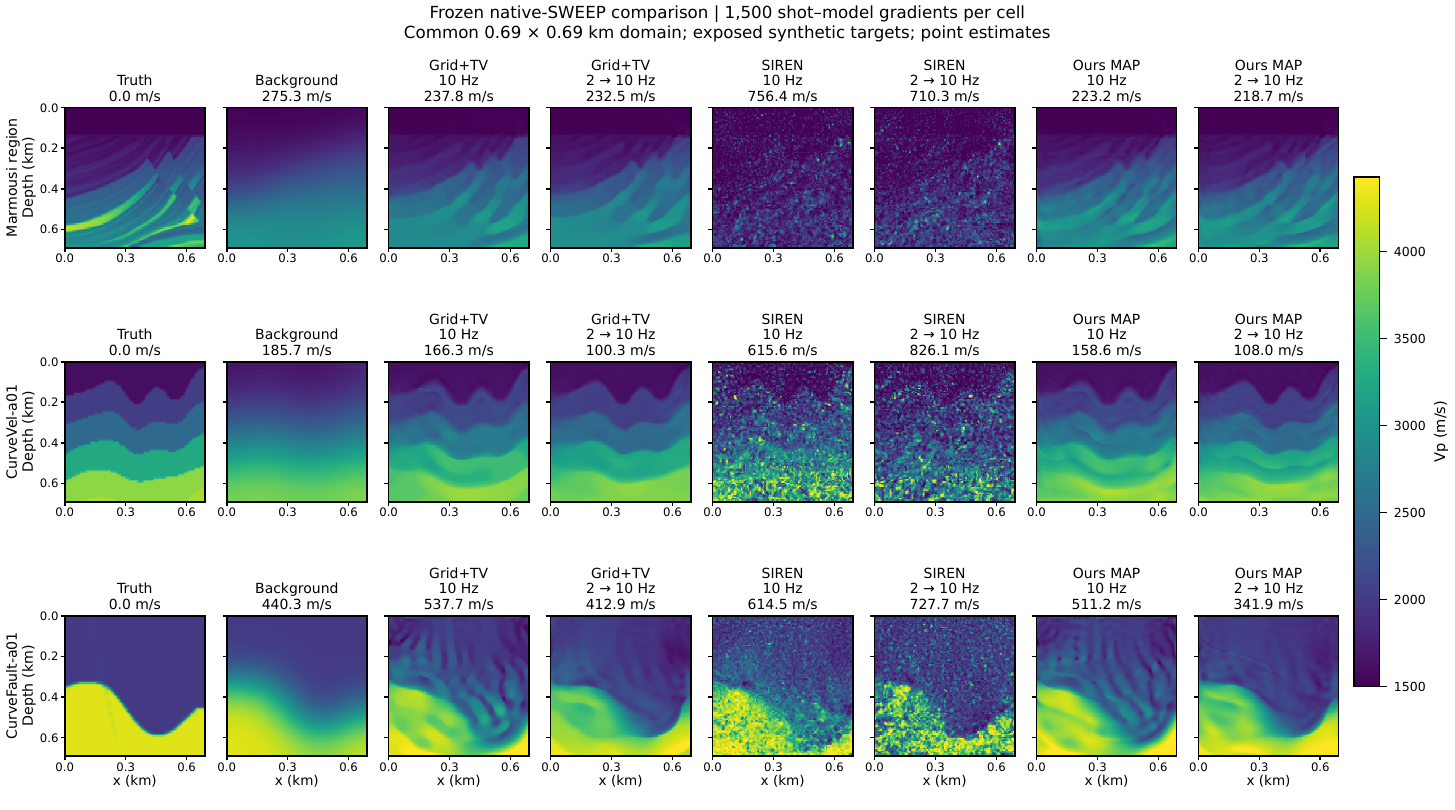}
\caption{\textbf{Frozen native-SWEEP comparison across three exposed targets.} Rows show rescaled Marmousi, CurveVel-a01, and CurveFault-a01; columns show truth, common background, and the three representations with direct 10 Hz and $2\rightarrow10$ Hz inversion. Every inversion uses 1,500 shot--model gradients. All coordinates refer to the common 0.69 km square computational domain. The velocity colour scale is the global truth envelope, shared by every panel; predictions outside this envelope are visually saturated, not numerically clipped. Full-range views and errors appear in Appendix~\ref{app:fields}. RMSE annotations use untruncated fields. The proposed model improves recovery on two targets, while continued Grid+TV is best on CurveVel. All SIREN endpoints remain worse than their backgrounds.}
\label{fig:main}
\end{figure}
\end{landscape}

Figure~\ref{fig:main} shows why an aggregate winner is an inadequate description. On Marmousi, Grid+TV and Ours recover inclined layering, with Ours reducing overall amplitude and position errors. Neither fully recovers the high-speed thin layers: even the best Ours endpoint reaches only about 3,482 m/s compared with a truth maximum of 4,429 m/s. On CurveVel, continued Grid+TV follows the broad layer undulations most closely. Ours is competitive but retains local bands and amplitude errors. SIREN produces dense granular variations and substantial structural error on all three targets.

The strongest schedule dependence occurs on CurveFault. Direct Grid+TV and direct Ours both produce prominent stripes and both degrade the background: their RMSEs are 537.67 and 511.23 m/s, corresponding to $\Delta_{\rm bg}=+22.10\%$ and $+16.10\%$. Continuation reduces these values to 412.86 and 341.86 m/s. Ours better approximates the main interface position and velocity contrast, but transition broadening and deeper errors remain. The selected truth does not justify describing this as unique recovery of fault topology, nor are stripes exclusive to the TV method.

\clearpage

\begin{figure}[tb]
\centering
\includegraphics[width=\linewidth]{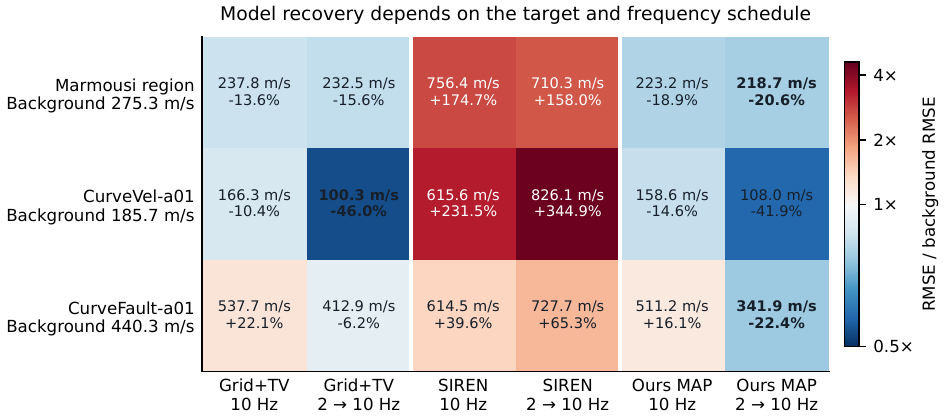}
\caption{\textbf{The ranking changes with structure and schedule.} Each cell gives velocity RMSE and its percentage change from the same target's background. Blue denotes improvement and red degradation; colour encodes the logarithm of the RMSE ratio. Bold entries mark each row's lowest RMSE. Normalizing to the background makes the comparison interpretable across targets with different initial errors, without treating them as independent samples of a population.}
\label{fig:heatmap}
\end{figure}

\subsection{Waveform prediction and velocity recovery disagree}
\begin{figure}[tb]
\centering
\includegraphics[width=\linewidth]{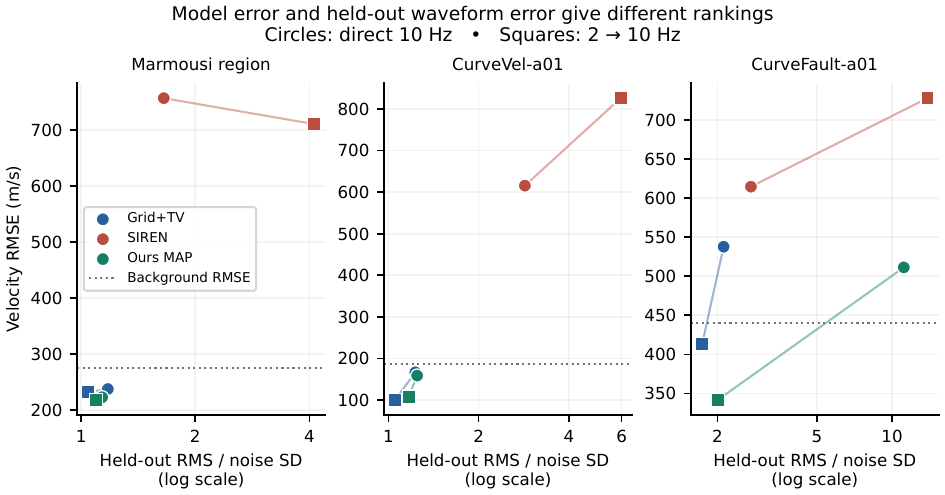}
\caption{\textbf{Better velocity recovery need not mean better unseen-shot prediction.} Each point is a final iterate; lower is better on both axes. Circles denote direct 10 Hz inversion and squares denote continuation, with connecting lines indicating the paired schedules. Horizontal dotted lines show background model RMSE, not background waveform scores. Held-out waveform residuals are normalized by the corresponding noise SD and displayed on a log axis. Grid+TV continuation predicts the unused shots better than Ours continuation on all three targets, although its model RMSE is worse on two.}
\label{fig:prediction}
\end{figure}

Figure~\ref{fig:prediction} and Appendix Table~\ref{tab:all} report an important reversal. Continued Grid+TV has lower held-out 10 Hz residuals than continued Ours on \emph{every} target: 1.045 versus 1.095 on Marmousi, 1.054 versus 1.170 on CurveVel, and 1.735 versus 2.009 on CurveFault, in noise-SD units. Ours is worse by 4.75\%, 10.97\%, and 15.81\%, respectively, on this predictive score. The two model-RMSE improvements therefore do not establish overall predictive superiority.

Direct Ours on CurveFault gives an especially clear warning: its training residual is 1.18, but its held-out residual is 11.14. Its model RMSE is lower than direct Grid+TV yet still worse than background. A fit concentrated on the optimized shot positions can coexist with an incorrect velocity field and poor transfer to unused shots. Model null spaces, uneven illumination, and finite-budget optimization are plausible contributors; this experiment does not isolate their individual effects.

\subsection{When frequency continuation worsens SIREN recovery}
\label{sec:siren_results}
All six SIREN endpoints are worse than their common backgrounds, but continuation has different effects across targets. It reduces Marmousi RMSE from 756.38 to 710.28 m/s (6.09\%), while increasing CurveVel from 615.62 to 826.15 m/s (34.20\%) and CurveFault from 614.51 to 727.69 m/s (18.42\%). Thus the defensible finding is a representation--schedule interaction, not a universal failure of frequency continuation.

\begin{figure}[tb]
\centering
\includegraphics[width=\linewidth]{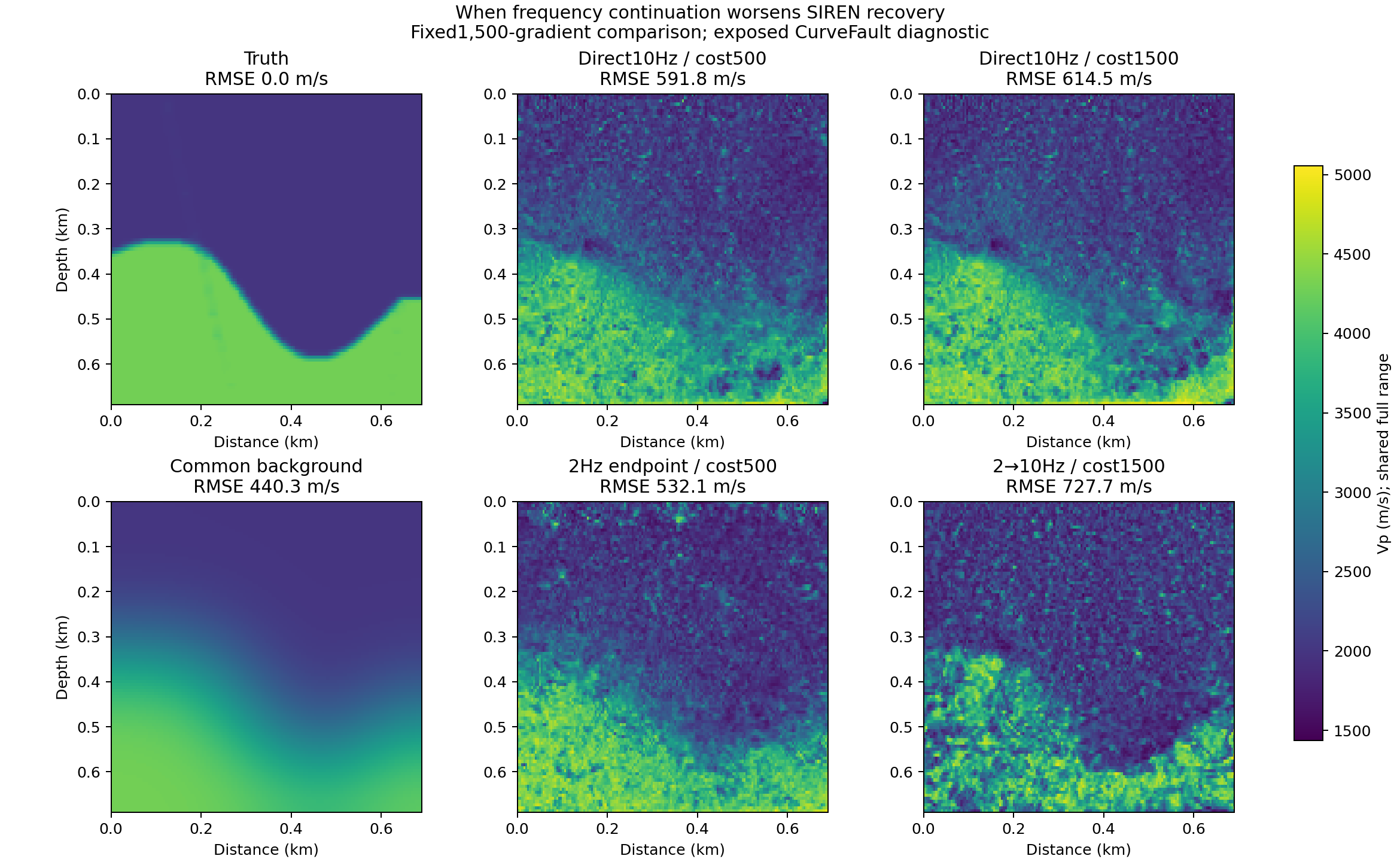}
\caption{\textbf{A conditional SIREN continuation failure on CurveFault.} Truth, background, stage-boundary fields, and final fields use the archived diagnostic layout. The continued run first improves after its 500-gradient switch and then deteriorates; the final field remains strongly granular. Endpoints are selected by budget, not by the best truth RMSE along the trajectory. Physical-gradient and absolute spectral-error measurements support increased final roughness, rather than an incorrectly smooth endpoint. Complete curves and spectra are in Appendix~\ref{app:spectral}.}
\label{fig:siren}
\end{figure}

The CurveFault trajectory helps constrain interpretation (Fig.~\ref{fig:siren}). At 500 gradients, the continued SIREN has RMSE 532.14 m/s, compared with 591.79 m/s for direct inversion at the same cost. After switching to 10 Hz, it briefly improves to a saved RMSE of 492.33 m/s at cost 599 before degrading to 727.69 m/s. A claim that it simply enters an incorrect smooth basin and then worsens monotonically is inconsistent with this trajectory. Training-shot residuals improve between the direct and continued endpoints (1.52 to 1.32), while held-out error rises from 2.72 to 13.86.

Spatial roughness also contradicts a ``smoother but more wrong'' description. The physical spatial-gradient RMS rises from 82.36 s$^{-1}$ for direct inversion to 110.32 s$^{-1}$ for continuation, a 33.94\% increase, whereas the truth value is 21.28 s$^{-1}$. Although the \emph{fraction} of spectral power above a declared high-wavenumber threshold decreases from 0.419 to 0.361, absolute high-band error RMS increases from 211.03 to 272.50 m/s. A smaller spectral fraction is not evidence of a smoother or more accurate field. Appendix~\ref{app:spectral} specifies the descriptive spectrum calculation.

Spectral bias is a useful framework for investigating neural optimization \citep{rahaman2019}, but these measurements do not demonstrate resonance between sine activations and gradient noise. The schedule changes the trajectory, the Adam state, and the amount of 10 Hz optimization (1,500 versus 1,000 gradients). None is independently controlled here. Moreover, SIREN was introduced specifically to represent detailed signals \citep{sitzmann2020}; one adapted configuration failing under a tight budget is not evidence of a universal representational inability.

\subsection{TV sensitivity does not reverse the CurveFault ranking}
\begin{figure}[tb]
\centering
\includegraphics[width=.95\linewidth]{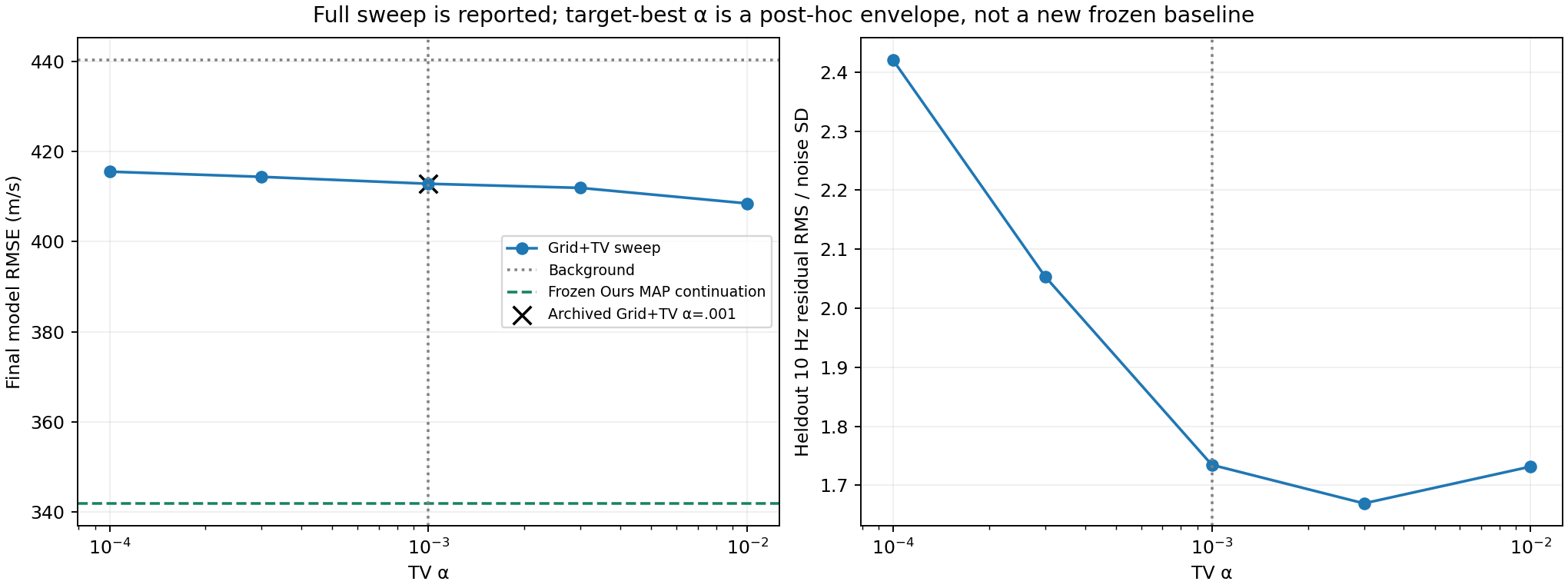}
\caption{\textbf{Post-exposure TV sensitivity on CurveFault.} All five weights use the same observations, background, optimizer, and 1,500-gradient continuation budget. The primary table retains the development-selected $\alpha=10^{-3}$. The smallest model error in this sweep occurs at its upper boundary, so the envelope is not a proven optimum. Model RMSE and held-out waveform error are reported separately.}
\label{fig:tv}
\end{figure}

Under the same 500+1,000 continuation budget, increasing the TV weight through $\{10^{-4},3\times10^{-4},10^{-3},3\times10^{-3},10^{-2}\}$ produces RMSEs of 415.55, 414.39, 412.86, 411.94, and 408.49 m/s (Fig.~\ref{fig:tv}). The repeated $10^{-3}$ run reproduces the archived baseline field exactly. The best scanned model RMSE improves on the frozen baseline by only 1.06\%, and Ours at 341.86 m/s remains 16.31\% lower than that sensitivity envelope.

This supports robustness of the observed ranking within the declared sweep. It does not establish a global Grid+TV optimum: the best model RMSE lies at the largest tested weight, some trajectories are still improving at the budget limit, and other step sizes or schedules were not searched. The best held-out waveform score in this sweep is 1.669 at $3\times10^{-3}$, again illustrating a distinction between model and predictive selection.

\subsection{Residual attribution and a negative variational result}
The saved components of Eq.~\eqref{eq:ours} challenge a simple attribution story. Under continuation, generator-increment/residual RMS amplitudes are 11.54/141.02 m/s on Marmousi, 17.47/146.26 m/s on CurveVel, and 100.60/240.89 m/s on CurveFault. In the first two cases, much of the visible layering lies in the residual component. These amplitudes are not additive variance fractions because the components can be correlated, but they show that the residual is an active reconstruction channel. The primary comparison cannot establish that a frozen generator supplies uniquely recoverable high-wavenumber structure.

The VI-8 extension on CurveFault is numerically finite but unsuccessful under the same PDE budget. Its posterior-mean RMSE is 469.22 m/s, worse than both the background (440.34) and MAP (341.86). The mean pointwise velocity SD is only 8.29 m/s, and nominal 90\% intervals cover 8.26\% of the true grid values. This is a severe empirical coverage failure for this target, not a calibrated posterior. It is also not clear evidence of variance collapse: mean field SD changes from 8.07 to 8.29 m/s, while mean latent-coordinate SD changes from 0.0200 to 0.01976. The approximation stays near a narrow initialization. The 63-update VI budget, initialization, and mean-field family have not been disentangled. Appendix~\ref{app:vi} reports these outcomes rather than treating VI as untested future work.
\FloatBarrier

%% file: data/rmse_table.tex
\begin{adjustbox}{max width=\linewidth}\begin{tabular}{llrrr}
\toprule
Method & Frequency & Marmousi & CurveVel & CurveFault \\
\midrule
Grid+TV & 10 Hz & 237.8 & 166.3 & 537.7 \\
Grid+TV & $2\rightarrow10$ Hz & 232.5 & \textbf{100.3} & 412.9 \\
SIREN & 10 Hz & 756.4 & 615.6 & 614.5 \\
SIREN & $2\rightarrow10$ Hz & 710.3 & 826.1 & 727.7 \\
Ours MAP & 10 Hz & 223.2 & 158.6 & 511.2 \\
Ours MAP & $2\rightarrow10$ Hz & \textbf{218.7} & 108.0 & \textbf{341.9} \\
\midrule
Background & --- & 275.3 & 185.7 & 440.3 \\
\bottomrule
\end{tabular}\end{adjustbox}

%% file: sections/discussion.tex
\section{Discussion}
\subsection{A conditional map of performance, rather than a universal winner}
The clearest supported result is that the best model-recovery method changes across the selected structures. Broad curved layering favours continued Grid+TV in this protocol. The rescaled inclined-layer and high-contrast-interface examples favour the generative-residual system in final velocity RMSE. These differences motivate structural diagnostics as part of FWI evaluation, but ``structure-dependent'' here describes three observed cases. It is not a causal isolation of geometry from contrast, scale, illumination, or background quality, nor an estimate over the full OpenFWI families.

A practical implication is to keep a properly regularized grid method in any neural comparison. On the CurveVel example, adding a pretrained generator does not improve the best grid result. On the more difficult CurveFault example, continuation is valuable for both grid and generative-residual inversion, and a TV-weight sweep does not erase their model-error gap. The adapted SIREN needs a different optimization or regularization study before it can be recommended under this tight budget. This conclusion applies to the tested implementation, not to all coordinate MLPs or all neural FWI formulations.

The physical intuition is that a parameterization changes the path by which waveform gradients can alter interfaces and amplitudes. A full-grid residual offers local freedom when a latent generator cannot express a needed correction. TV instead penalizes the aggregate size of spatial changes. Neither observation implies that TV cannot preserve edges or that the generator necessarily carries the fine detail. Indeed, our component fields show substantial residual participation. A useful next comparison would hold residual scaling, Gaussian penalty, clipping, and schedule fixed while switching between prior-only, residual-only, and joint models; a random-decoder control would further distinguish pretrained content from architectural conditioning.

\subsection{Continuation is an optimization intervention}
Multiscale waveform inversion is intended to improve access to useful solutions \citep{bunks1995,virieux2009}. Our grid and generative-residual endpoints improve with continuation on all three cases, whereas the SIREN response is mixed. Since both model state and Adam moments carry across the switch, the experiment tests a complete continuation procedure. It does not identify whether the adverse SIREN outcomes originate in a low-frequency model, optimizer memory, a reduced high-frequency budget, or their interaction.

Several additional controls would be needed to turn the present spectral observations into a mechanism: restart versus retain Adam moments at the switch, allocate equal high-frequency cost as a separately labelled larger-budget test, vary the sinusoidal frequency parameter on development data, and compare representation-gradient spectra with final model spectra. These are future experiments, not analyses silently inferred from the current panels. The existing generator already uses bilinear resize-plus-convolution upsampling (Appendix~\ref{app:prior}), so a transposed-convolution overlap explanation does not apply to this checkpoint.

\subsection{Waveform agreement, geometry, and uncertainty are different tests}
The disparity between velocity RMSE and held-out prediction is scientifically consequential. A velocity field that is closer in a global pixelwise norm can still make worse predictions on source positions not used for fitting. Conversely, lower held-out waveform error is not a proof of a uniquely correct velocity field. Both should be retained, along with structural diagnostics appropriate to the target, rather than selecting whichever supports a preferred method.

This perspective is consistent with the broader \emph{Relational Geophysics} programme motivating the work: physically useful inference should respect relationships among observations and the geometrical organization of the inferred field, not only local fit. Here that is a research perspective, not an additional validated theorem. The disagreement between training and unused-shot predictions is a concrete reason to test those relationships. No evidence from separate focal-mechanism or signal-quality studies is imported into the FWI claims.

Similarly, a successful MAP reconstruction does not imply that an associated variational posterior is informative. VI approximates a distribution through optimization within a chosen family \citep{blei2017}. In our equal-cost extension, its mean field underperforms the background and its intervals fail the available coverage check. Reporting this result limits the present contribution to point-estimate comparison and diagnostics; it provides no evidence of variational superiority or calibrated uncertainty.

\subsection{Limitations and the next benchmark}
The study is restricted to two-dimensional constant-density acoustics, known source signatures, synthetic Gaussian noise, one numerical operator, one seed, and one exposed target per family. Truth-smoothed backgrounds provide information that would need to be estimated in a field application. The Marmousi remapping changes its physical length scales and aspect ratio. Sparse held-out sources test prediction within the same synthetic acquisition, not transfer to new geological targets, field data, or another solver.

A fixed operator isolates numerical choices across representations, but synthesis and inversion using the same discretization is an inverse-crime setting. Development refinements change traces by about 5.10\% at 2 Hz and 8.54\% at 10 Hz; these checks are not a full grid-convergence study. The 1.8 s record and the low-frequency source delay are also specific to this small domain. Native-scale and independently discretized tests remain necessary.

Computational matching is deliberately limited to PDE-gradient cost. SIREN has many more optimized weights than either grid or generative-residual inversion, and VI spends gradients on multiple samples rather than additional parameter updates. Pretraining and background fits are extra costs. The main TV search is short, the sensitivity optimum is not bracketed, and the systems use different priors, step sizes, clipping, and output constraints. We therefore make no claim about asymptotic optimality or a representation-only causal effect.

A stronger subsequent benchmark should freeze all selection rules on separate development targets, use multiple unseen targets and seeds in each family, include alternative acquisition and modelling errors, and publish failures alongside successes. Structural metrics should be declared before examining outputs; in particular, a family named CurveFault should not automatically be scored as successful fault-topology recovery. The current archive is a diagnostic foundation for that protocol, not a substitute for it.

%% file: sections/conclusions.tex
\section{Conclusions}
Using a common SWEEP acoustic discretization and 1,500 shot--model gradients per cell, we find conditional, rather than universal, advantages among three FWI representations. With continuation, the frozen-generator/full-grid-residual system achieves the lowest velocity RMSE on rescaled Marmousi and the selected CurveFault example, whereas Grid+TV is best on CurveVel. Grid+TV predicts held-out 10 Hz shots better on all three. The tested SIREN configuration is worse than background throughout, but continuation harms only two of the three targets, and its mechanism remains unresolved. The CurveFault model-error ranking survives the declared TV sweep, while an equal-budget VI extension fails recovery and coverage.

These findings support method auditing over a single superiority claim. A defensible neural FWI comparison must report its actual degrees of freedom, common initialization, complete computational budget, residual regularization, target provenance, model errors, predictive errors, and negative outcomes. The next step is a preregistered multi-target evaluation and a matched attribution experiment that separates the generator's contribution from full-grid residual freedom.

%% file: sections/appendix.tex
\section{Frozen generator and reproducibility details}
\label{app:prior}
The scalar VAE has six encoder blocks with channels $[24,48,96,128,192,256]$. Each uses a $4\times4$ stride-two convolution, group normalization, and SiLU activation. At $256\times256$ input resolution, the final feature map has $256\times4\times4$ entries. Separate linear heads predict the 128-dimensional latent mean and log variance. The decoder maps the latent to that feature shape and upsamples through channels $[256,192,128,96,64,40,24]$. Each decoder block uses bilinear interpolation by a factor of two (unaligned corners), a $3\times3$ convolution, group normalization, and SiLU. A final $3\times3$ single-channel convolution and sigmoid yield a native $256\times256$ image. This is resize-plus-convolution, not transposed-convolution upsampling. The complete VAE has 3,911,217 parameters; its weights are frozen for all FWI runs.

The image corpus comprises 27,000 EuroSAT images, 500 BSDS500 images, and 5,640 DTD images. The manifest assigns 29,816 identities to training and 2,000 to validation; the remaining 1,324 identities are unused. Source-balanced draws use probabilities 0.45, 0.20, and 0.35, respectively. Preprocessing includes grayscale conversion, 1st--99th percentile normalization, random cropping, right-angle rotations, reflections, contrast inversion, and mixtures of fine and coarse Gaussian smoothing. The training seed is 20260814, with 30 epochs of 30,000 augmented draws, batch size eight, and AdamW learning rate $3\times10^{-4}$ with weight decay $10^{-5}$ and cosine learning-rate decay to $10^{-6}$. The loss combines image MSE, gradient $L^1$ error with weight 0.1, and KL regularization ramped over eight epochs to $5\times10^{-5}$, with a 0.02 per-coordinate free-bits floor. The archived best epoch is 29, chosen by image validation error. No FWI target selected this checkpoint.

The inference checkpoint SHA-256 is\par
\noindent\texttt{37d9494a4b9ccffb6f3c7ca31aefc207576dcbc6a03a28ec8ae814da6b2daf13}.\par
The archived SWEEP notebook checkout is identified by commit\par
\noindent\texttt{013640b28f2b6b1ad0b2ffd701c7253f05dc80a4}.\par
Package source hashes, not just this notebook checkout, identify the numerical engine. The manuscript bundle includes numerical tables, figure checksums, and machine-readable method summaries. It does not embed the large checkpoint or raw waveform arrays.

The background latent fit starts from the VAE encoder mean of the resized normalized background, $\operatorname{clip}[(R_{256}m_0-1500)/3000,0,1]$. It uses Adam for 500 steps, learning rate 0.04 with cosine decay to 0.002, and an image MSE plus $10^{-5}$ latent mean-square penalty. SIREN receives 500 background-only Adam steps at learning rate 0.001, fitting the bounded physical field before anchoring. These initializations carry background information into the parameterization without changing the exactly common initial physical model.

\section{Complete endpoint metrics and gradient trajectories}
\label{app:metrics}
\begin{figure}[p]
\centering
\includegraphics[width=\linewidth]{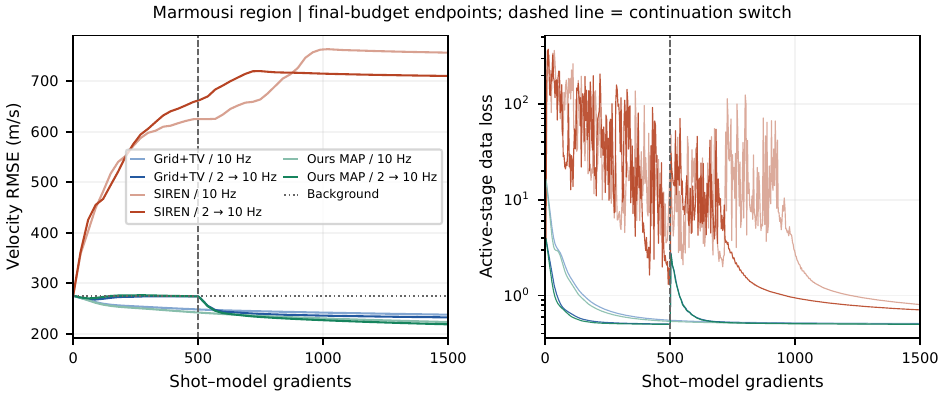}\par\vspace{2mm}
\includegraphics[width=\linewidth]{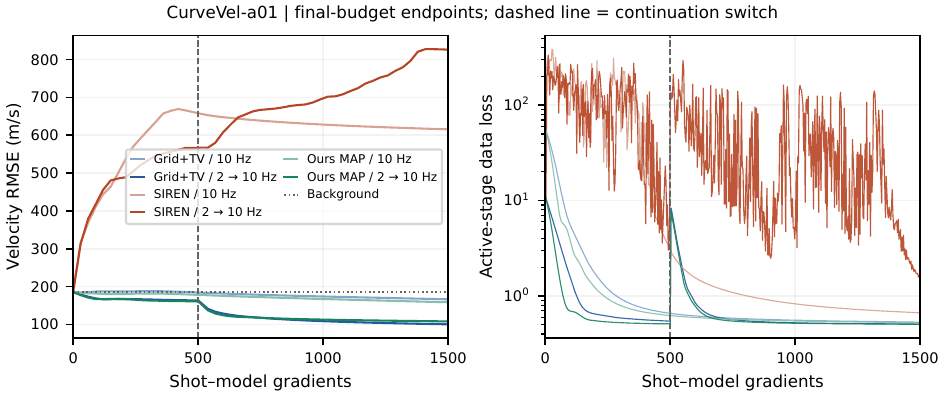}\par\vspace{2mm}
\includegraphics[width=\linewidth]{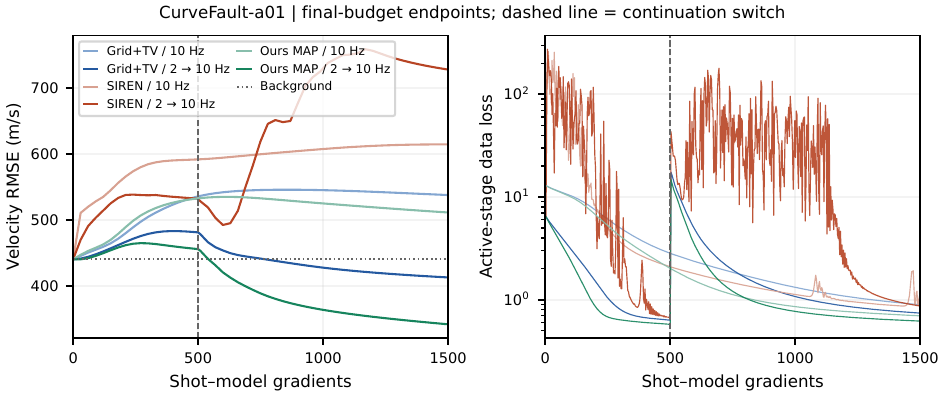}
\caption{\textbf{Complete gradient--error trajectories.} Rows correspond to Marmousi, CurveVel, and CurveFault. Left: truth RMSE scored from saved fields after optimization. Right: active-stage waveform loss logged during fitting. The vertical line at 500 marks a frequency change only for continued methods; direct runs remain at 10 Hz throughout. No intermediate RMSE minimum is used to choose the headline endpoint. Different objectives are active before and after the switch, and some methods are still improving at cost 1,500.}
\label{fig:curves}
\end{figure}

Table~\ref{tab:all} includes both negative and positive background-relative changes. All entries completed 1,500 gradients; all Ours guards remained inactive. A guard is not applicable to the grid and SIREN comparators. Every reported RMSE uses the final field, without colour-scale clipping. Figure~\ref{fig:curves} gives saved trajectories. The data loss is recorded before each Adam update, whereas saved field RMSE is scored after an update. Consequently points at the same charged-cost label need not correspond to precisely the same parameter snapshot. The active data loss also changes its source/noise slice at the continuation boundary; it is not a common held-out objective across that boundary.

\begin{table}[htbp]
\centering\small
\caption{All primary endpoints. Velocity RMSE is in m/s; $\Delta_{\rm bg}<0$ is improvement. Train and held-out columns are 10 Hz residual RMS divided by noise SD. Numerical completion is not evidence of successful recovery.}
\label{tab:all}
\input{data/all_metrics_table}
\end{table}
\clearpage
\clearpage

\begin{figure}[p]\centering
\includegraphics[width=.9\linewidth]{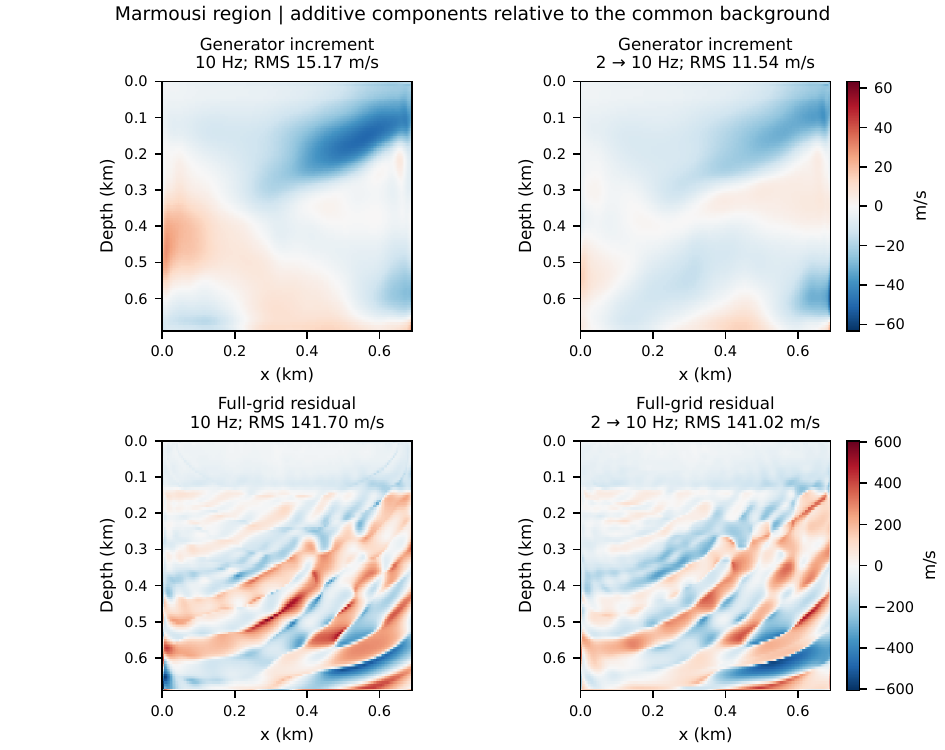}
\caption{Marmousi generator increments and full-grid residuals for both schedules. Under continuation their RMS amplitudes are 11.54 and 141.02 m/s. Layering is predominantly expressed by the residual; this comparison does not show that the generator supplies the missing high-wavenumber structure.}
\label{fig:compm}\end{figure}

\begin{figure}[p]\centering
\includegraphics[width=.9\linewidth]{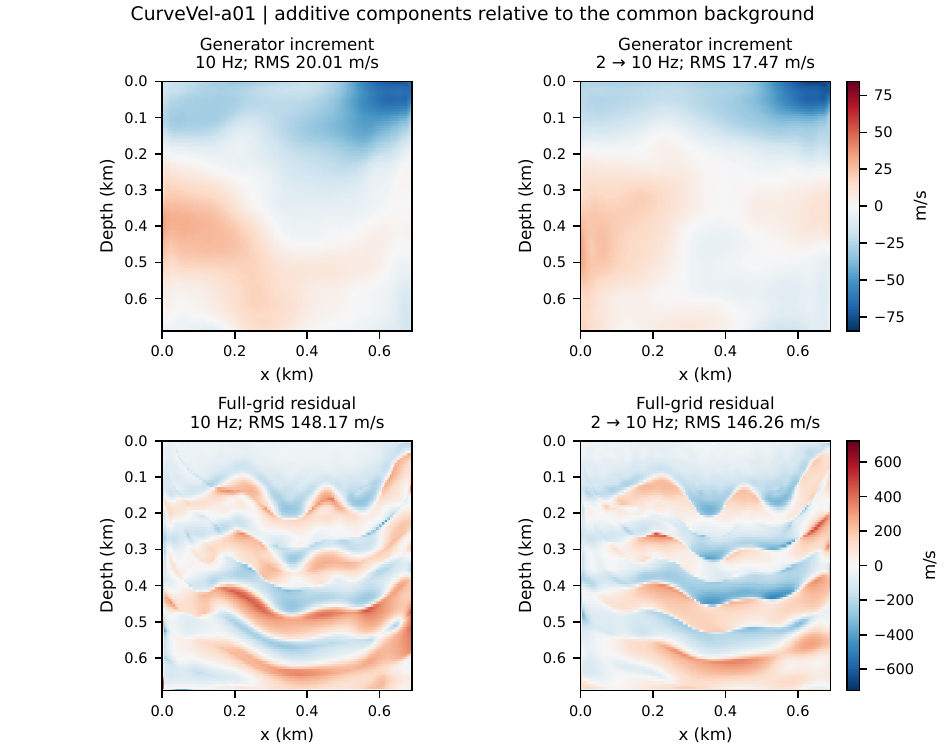}
\caption{CurveVel generator increments and residuals. Their continuation RMS amplitudes are 17.47 and 146.26 m/s. The generator increment is comparatively smooth; the residual contains substantial interface detail and local error.}
\label{fig:compv}\end{figure}

\begin{figure}[p]\centering
\includegraphics[width=.9\linewidth]{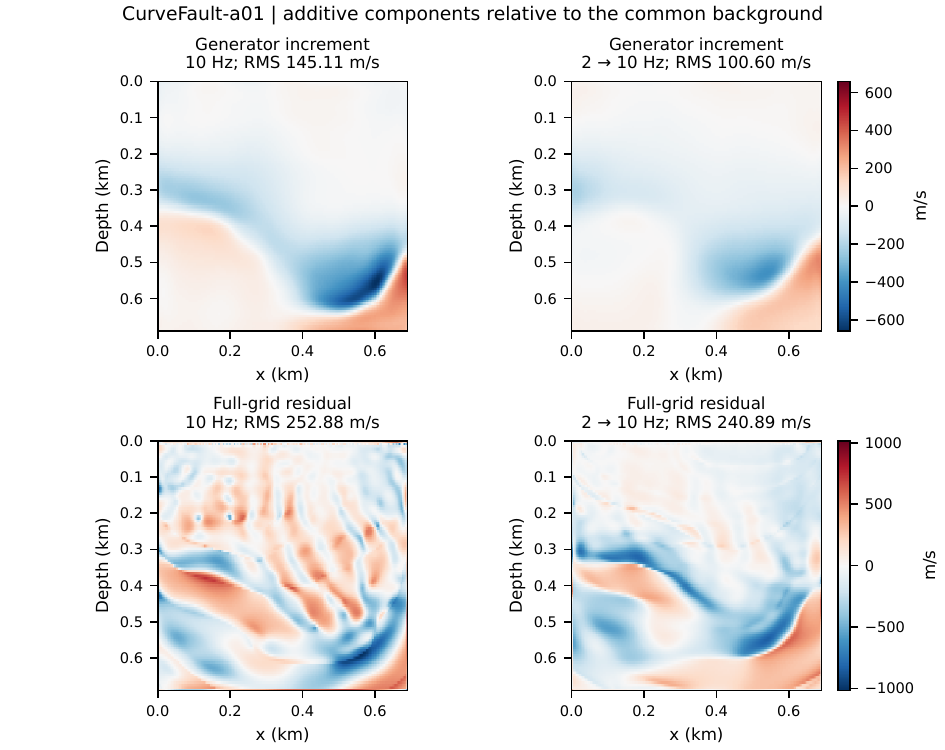}
\caption{CurveFault generator increments and residuals. Continuation RMS amplitudes are 100.60 and 240.89 m/s, larger generator participation than in the other examples. Correlation between components prevents interpreting these amplitudes as independent explained-variance fractions.}
\label{fig:compf}\end{figure}

\begin{landscape}
\section{Full-range fields and residual attribution}
\label{app:fields}
The common colour scale in Fig.~\ref{fig:main} aids structural comparison but can hide excursions outside the truth envelope. Figures~\ref{fig:fullm}--\ref{fig:fullf} therefore include full-range reconstructions and signed error maps. Figures~\ref{fig:compm}--\ref{fig:compf} expose both terms added to the background in Eq.~\eqref{eq:ours}. In particular, visible layers in the residual mean that the joint representation should not be described as a 128-degree-of-freedom inversion. Component magnitude alone is not a causal measure of contribution; opposing or correlated changes can occur because the decomposition is nonunique.

\begin{figure}[htbp]\centering
\includegraphics[width=\linewidth]{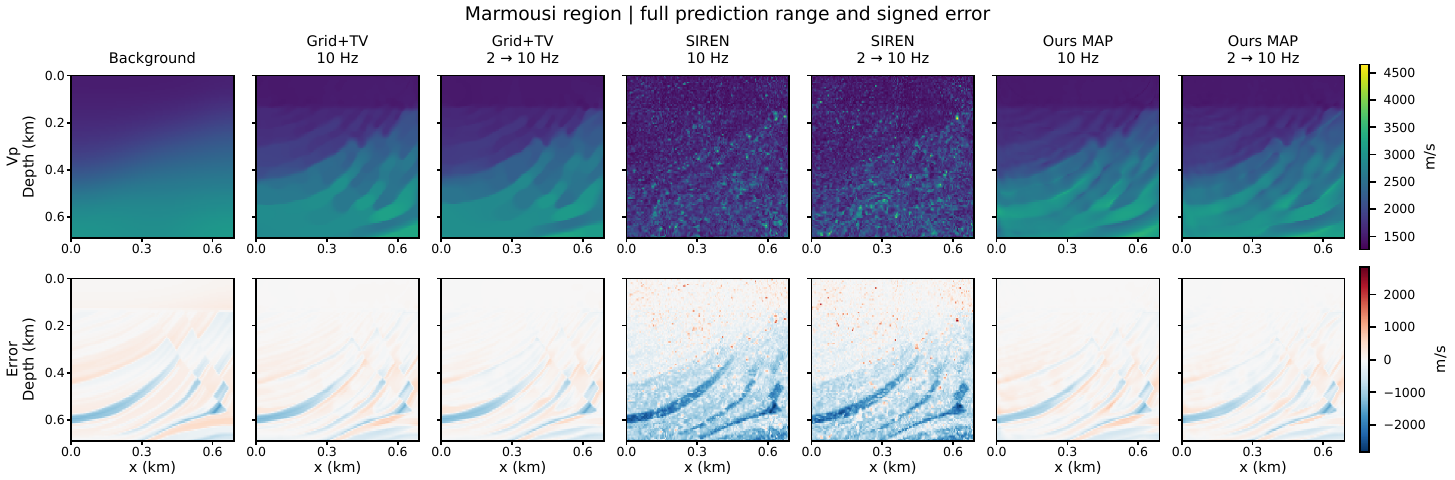}
\caption{Marmousi full-range fields and signed velocity errors. Predictions outside the global truth envelope remain visible. Inclined layers are recovered by grid and joint methods, but the fastest thin layers remain underestimated. The same rescaled computational geometry applies to every column.}
\label{fig:fullm}\end{figure}
\clearpage
\begin{figure}[p]\centering
\includegraphics[width=\linewidth]{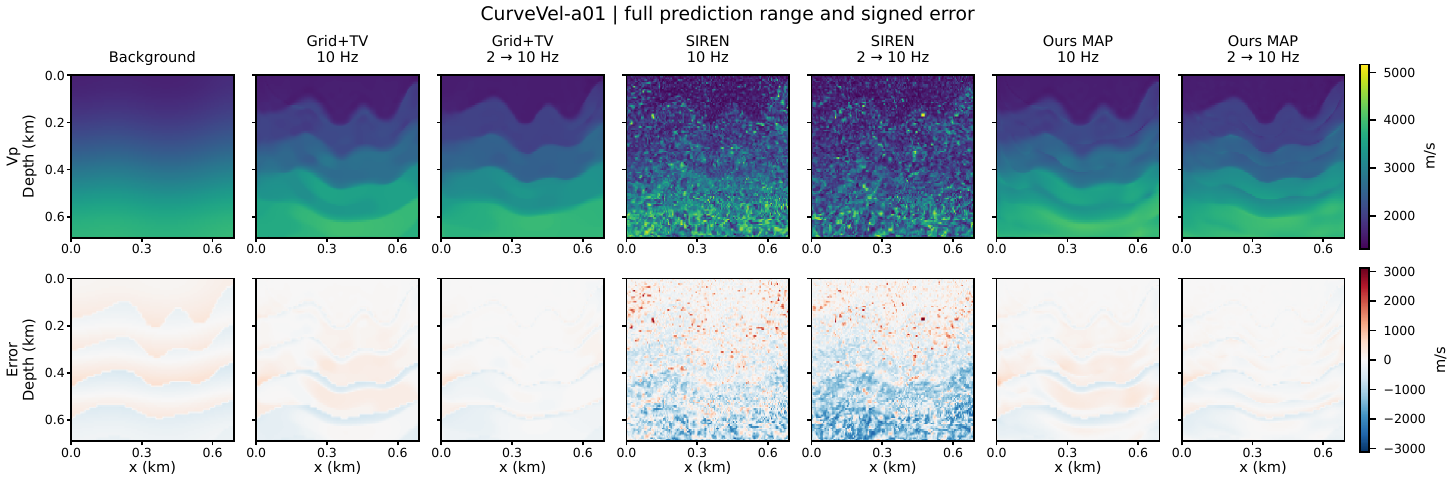}
\caption{CurveVel full-range fields and signed velocity errors. Continued Grid+TV gives the lowest global RMSE. Ours continuation is close but has residual bands and local amplitude bias. Granular SIREN errors extend throughout the section.}
\label{fig:fullv}\end{figure}
\clearpage
\begin{figure}[p]\centering
\includegraphics[width=\linewidth]{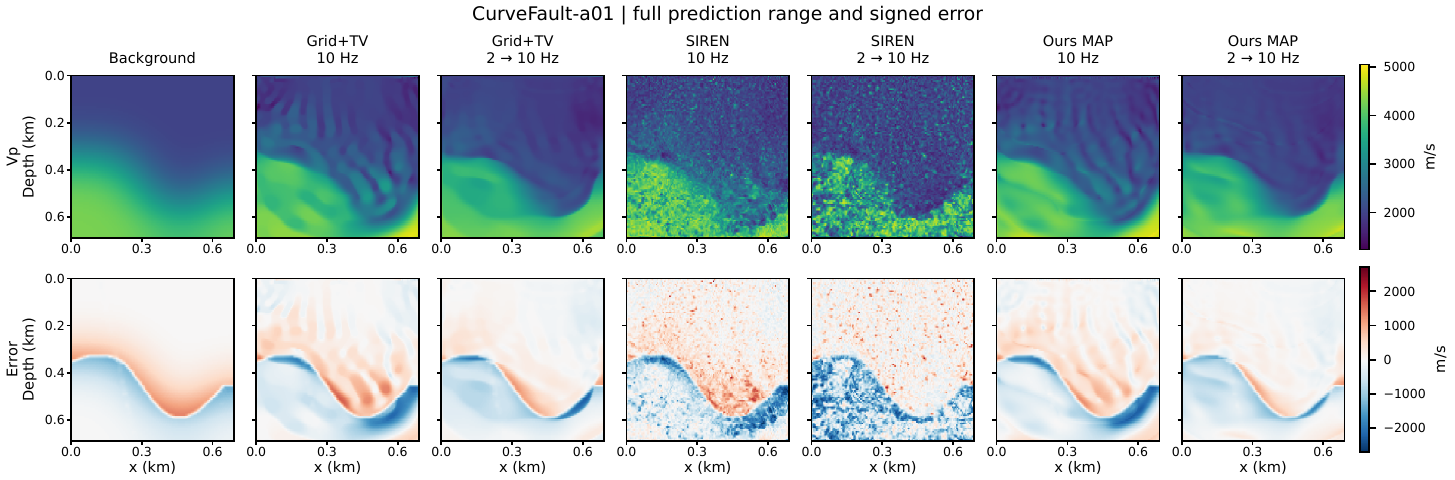}
\caption{CurveFault full-range fields and signed velocity errors. Direct Grid+TV and Ours both develop stripes and both underperform the common background. Continuation reduces the errors but does not completely recover the sharp true boundary. This sample contains a curved contrast boundary rather than an unambiguous fault-offset topology.}
\label{fig:fullf}\end{figure}
\end{landscape}
\clearpage

\section{SIREN spectrum and TV diagnostics}
\label{app:spectral}
\begin{figure}[htbp]\centering
\includegraphics[width=\linewidth]{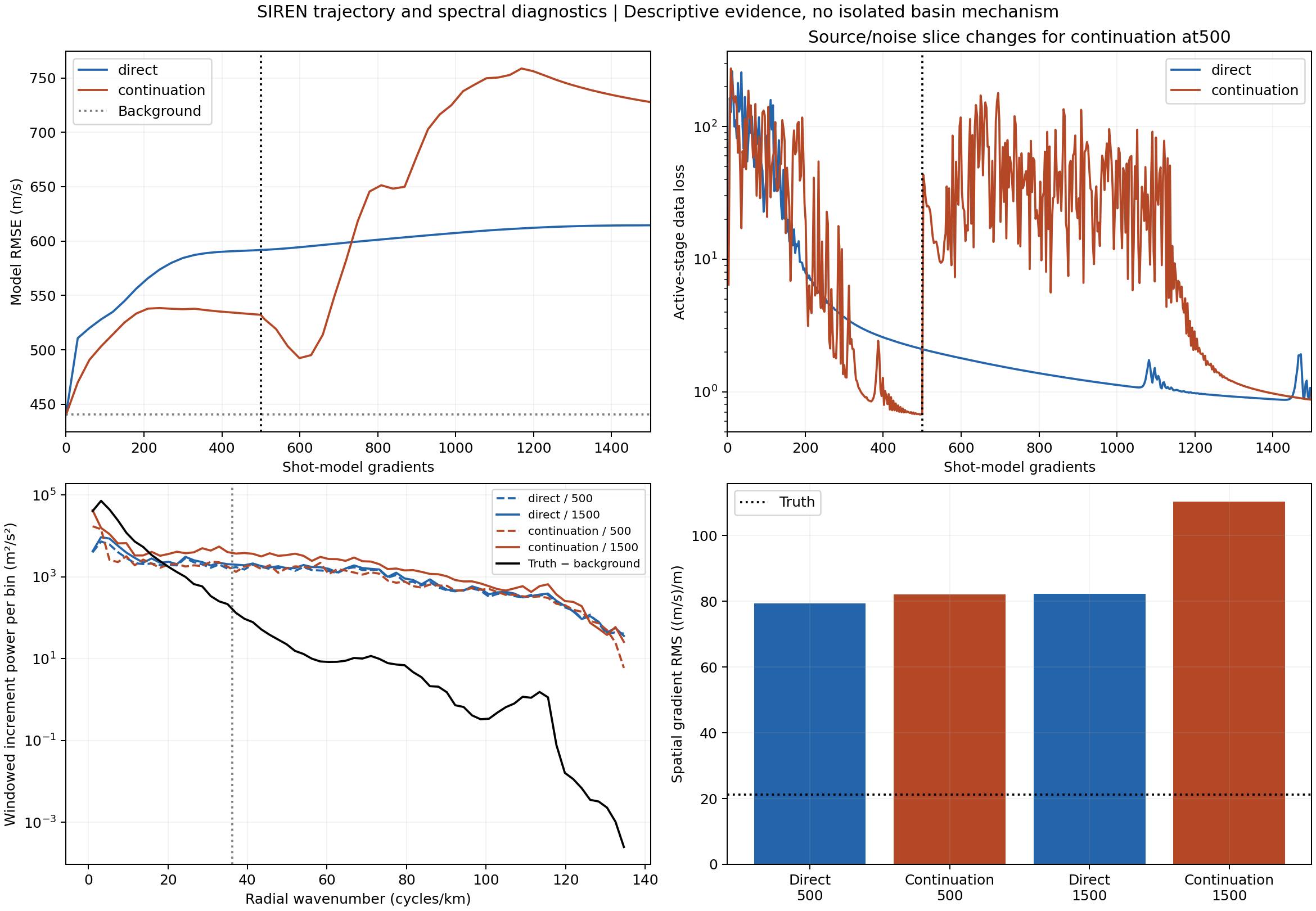}
\caption{SIREN CurveFault trajectories and wavenumber diagnostics. The continued run briefly improves after the switch before worsening. Its final absolute high-band error and spatial-gradient RMS exceed those of the direct run, although its high-band \emph{fraction} is lower. These measurements describe the failure; they do not isolate an activation-induced resonance mechanism.}
\label{fig:spectrum}\end{figure}

The descriptive spectrum analysis uses the model increment relative to the background. A weighted mean is removed, a two-dimensional Hann window is applied, and an orthonormal two-dimensional FFT is radially aggregated into 64 bins in physical cycles/km. The high-band cutoff is 36.2319 cycles/km, half of the smaller axial Nyquist frequency. This is a declared grid-scale descriptor, \emph{not} a theoretical FWI resolution limit. Absolute error spectra are computed separately from power fractions. Spatial-gradient RMS uses physical grid spacings. Figure~\ref{fig:spectrum} shows why a relative power fraction must be read together with absolute error and roughness.

\begin{landscape}
\begin{figure}[p]\centering
\includegraphics[width=\linewidth]{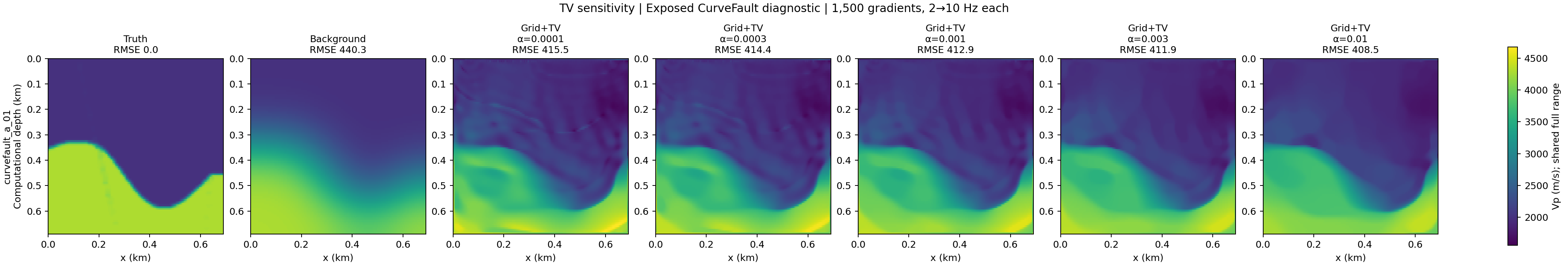}
\caption{All five CurveFault TV-weight outcomes under continuation. Each receives the same 1,500-gradient budget. The main result uses the independently development-selected $10^{-3}$; the other weights are a post-exposure sensitivity analysis. The largest tested weight has the lowest model RMSE, so further improvement outside the sweep has not been excluded.}
\label{fig:tvfields}\end{figure}
\end{landscape}

The TV sensitivity fields in Fig.~\ref{fig:tvfields} retain all five weights. Their held-out waveform residuals are 2.421, 2.054, 1.735, 1.669, and 1.732 in increasing-weight order. The $10^{-3}$ repetition is pointwise identical to the main baseline, providing an additional cross-run consistency check.

\section{Equal-budget variational inference and coverage failure}
\label{app:vi}
For the model in Eq.~\eqref{eq:ours}, define $w=(z,q)$ and a diagonal Gaussian $Q_\lambda(w)=\N(\mu,\operatorname{diag}(\exp(2\ell)))$. There are 32,778 optimized variational scalars. VI minimizes
\begin{equation}
 J_{\rm VI}=\E_{Q_\lambda}[\mathcal{L}_f(m(w))]
 +\frac{1}{N_{\rm obs}}\KL\left[Q_\lambda(w)\,\|\,\N(0,I)\right],
\end{equation}
using reparameterized Monte Carlo gradients \citep{kingma2014,blei2017}. The eight draws per full update are four independent antithetic pairs. Means use the MAP learning rates; both log-SD blocks use 0.002. Initial coordinate SDs are 0.02, corresponding to 6 m/s residual SD before adding generator variability. Each block is clipped at norm 10. No free initialization from the already optimized MAP is supplied.

A normal update costs $8\times3=24$ shot--model gradients. Smaller shot subsets retain all eight samples at budget boundaries, yielding 21 low-frequency and 42 high-frequency updates, with exactly 500 total gradients per training shot. The unequal update count relative to MAP (63 versus 501) is a consequence of equal PDE cost, not an unreported computational advantage. The final objective concerns the 10 Hz data; the 2 Hz likelihood is not retained as an extra term.

All sampled fields undergo the same operational physical-support checks as MAP. This does not define a normalized hard-truncated Gaussian posterior: Gaussian coordinates have unbounded tails, and passing a finite set of checks is not a proof of hard support or calibration. No variance clamping, resampling, or post hoc fallback is used.

We evaluate the physical posterior mean using 128 fixed independent draws, separately from the central field $m(\mu)$. At initialization their RMSEs are 440.41 and 440.34 m/s; finally they are 469.22 and 469.08 m/s. Pointwise empirical 5th--95th percentiles have mean width 26.69 m/s and cover 8.26\% of the truth. Coverage is a descriptive spatial fraction for one target, not a repeated-sampling calibration study. Nevertheless, its disagreement with nominal 90\% coverage is large.

\begin{figure}[p]\centering
\includegraphics[width=\linewidth]{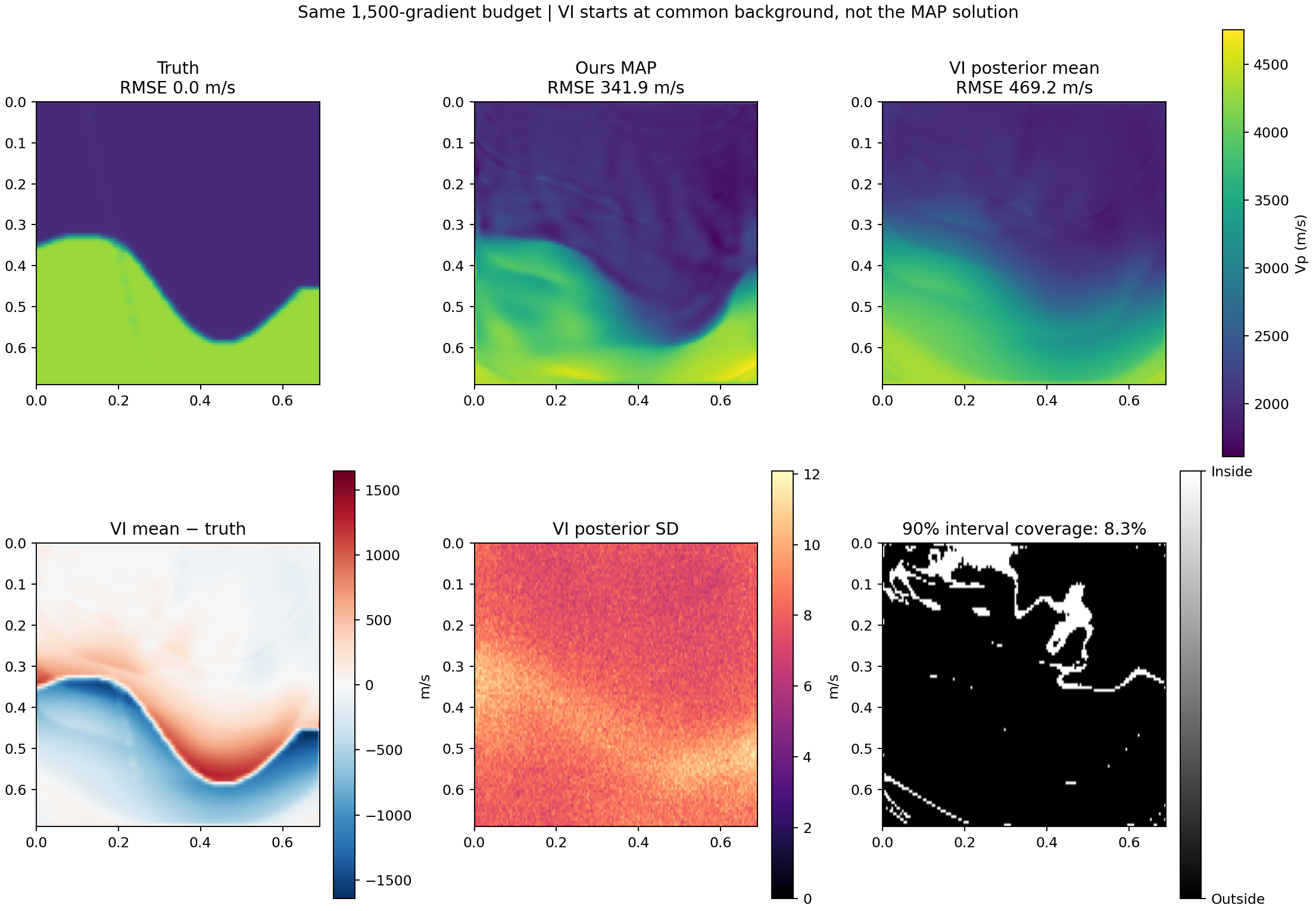}
\caption{CurveFault VI-8 under the same 1,500 shot--model gradient budget. Posterior mean, pointwise uncertainty, and error/coverage diagnostics are shown alongside the relevant references. Numerical completion does not establish accuracy: mean-field RMSE is 469.22 m/s and nominal 90\% intervals cover only 8.26\% of the true velocity values.}
\label{fig:vi}\end{figure}

\begin{figure}[p]\centering
\includegraphics[width=\linewidth]{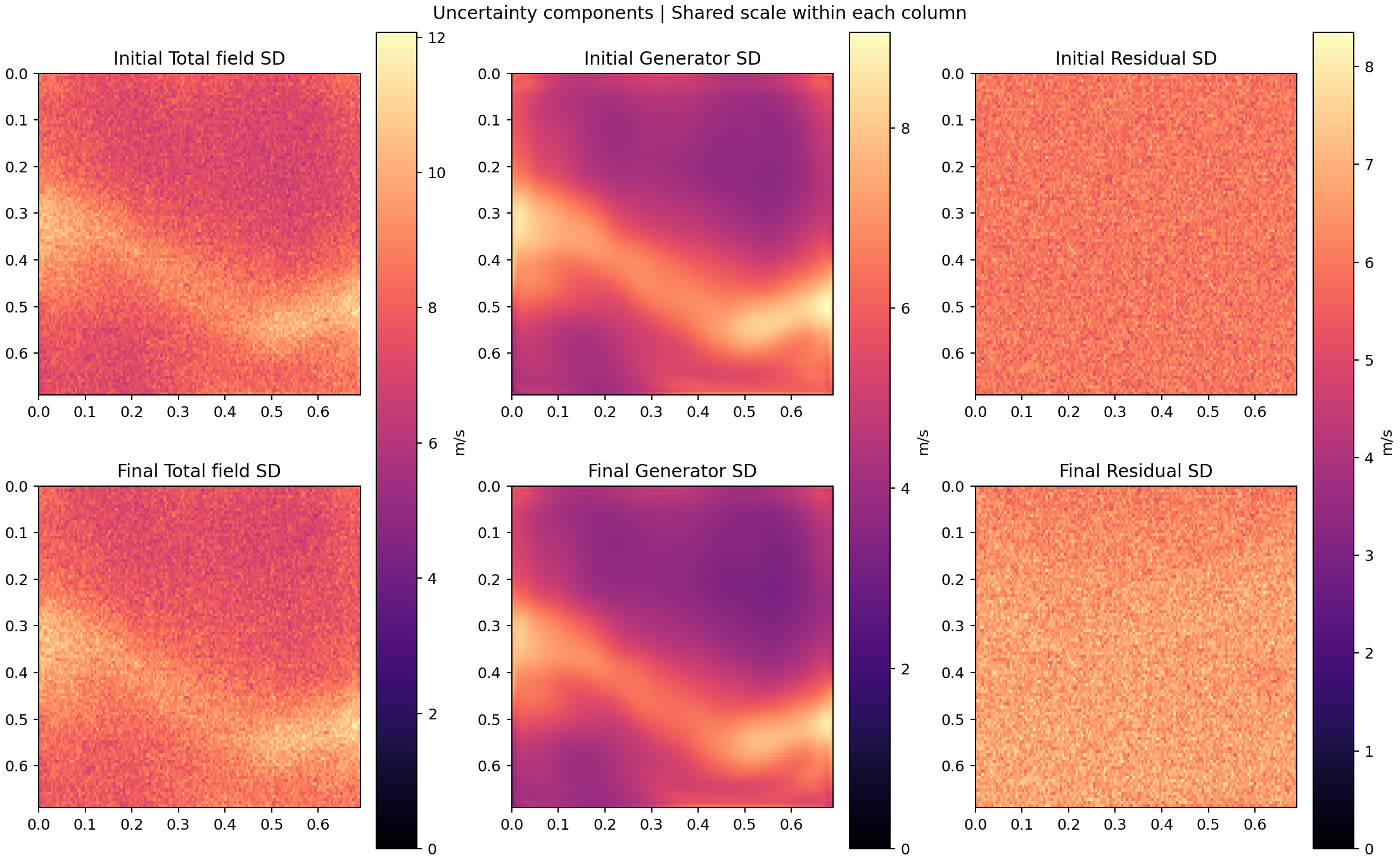}
\caption{Initial and final uncertainty components for VI-8. The approximation remains near a narrow initialization. Mean total SD increases slightly, so the endpoint cannot be diagnosed as a large variance collapse merely from its small absolute SD.}
\label{fig:visd}\end{figure}

\begin{figure}[p]\centering
\includegraphics[width=\linewidth]{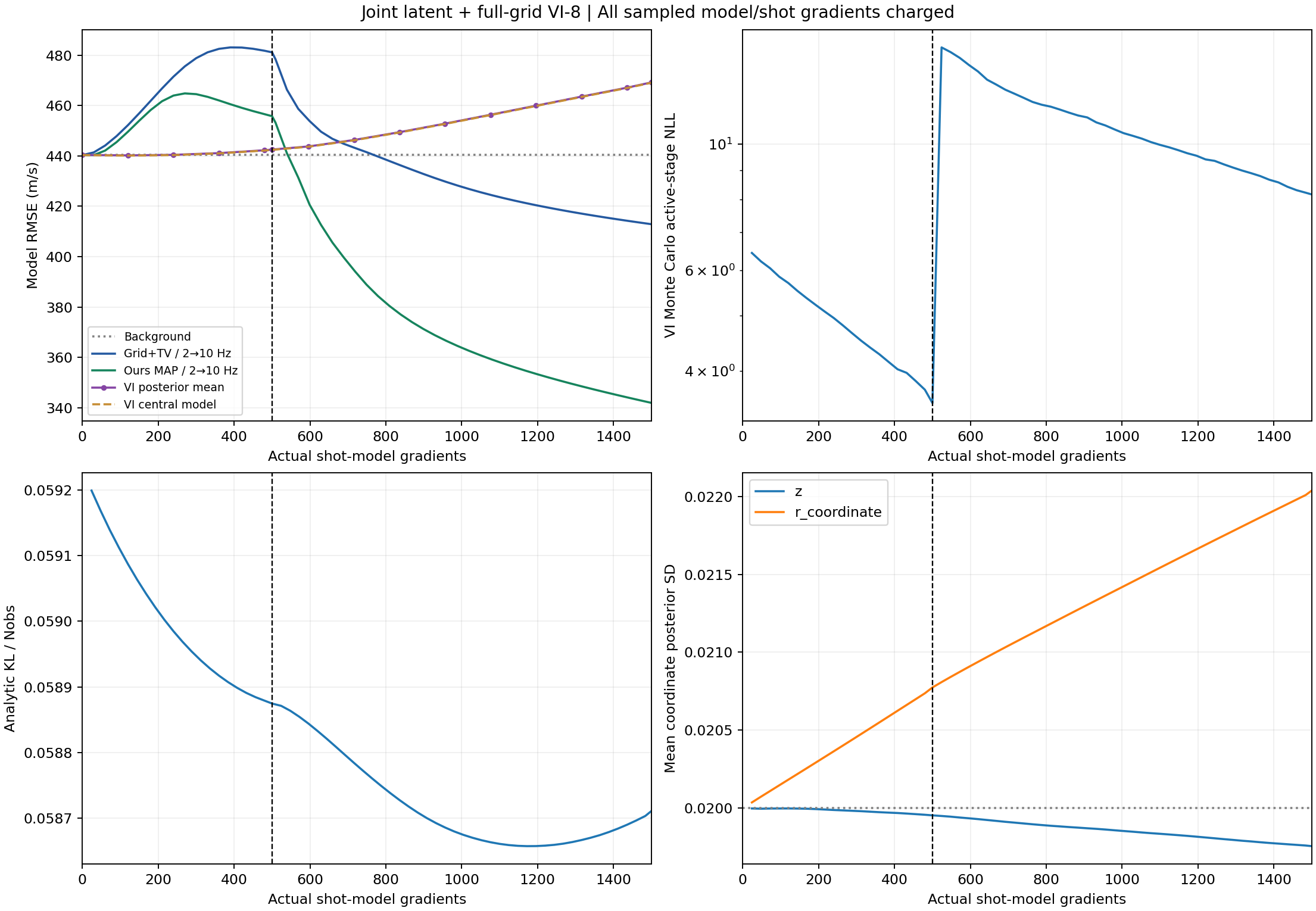}
\caption{VI-8 optimization and uncertainty traces against actual PDE-gradient cost. The strict budget permits only 63 updates, compared with 501 for MAP. Active-stage likelihood changes at cost 500. These curves do not demonstrate posterior convergence or isolate the effects of budget, initialization, and mean-field factorization.}
\label{fig:vitrace}\end{figure}

The latent-coordinate SD remains near its initial value (0.0200 to 0.01976), while residual-coordinate SD grows from 0.0200 to 0.02204. Mean total field SD rises from 8.07 to 8.29 m/s. Thus the observed failure is a narrow, inaccurate approximation, without a large measured collapse from its already narrow initialization. The mean-model held-out waveform residual is 4.392. A 32-draw predictive ensemble gives mean residual 4.402; nominal 90\% observed-trace coverage is 4.56\% using epistemic variability alone and 66.14\% after adding observation noise. Figures~\ref{fig:vi}--\ref{fig:vitrace} retain these negative results. An equal-optimizer-update, larger-PDE-budget VI experiment has not been run.

\clearpage

\section{Execution audits and scope of cost accounting}
The primary 18-cell comparison contains 27,000 optimization shot--model gradients. The additional five-point TV sweep costs 7,500, and VI-8 costs 1,500. The resulting 36,000 total excludes development selection, gradient audits, observation synthesis, forward-only diagnostics, and network pretraining/background fits. Historical results are reused, not charged as newly executed runs. In the completion stage, the new CurveVel matrix, TV sweep, and VI run together cost 18,000 gradients; previously archived Marmousi and CurveFault matrices account for the other 18,000.

The completion ledger separately records 124 final diagnostic forward shot--models, 10 new observation-synthesis forwards, and 98 forwards/50 adjoints for the VI development audit. CurveVel adds 1,000 background-only network gradient steps, with no PDE evaluation. VI field-draw diagnostics also involve decoder evaluations without wave propagation. These figures are stage-specific accounts, not a claim that all historical auxiliary work is included in the 36,000 optimization total.

Independent-wavefield grouping was checked against separate propagations. An initial short-record batching check exceeded its declared tolerance and is retained in the archive; the accepted full 4,500-sample audit subsequently found identical forward values and parameter gradients for the final grouping. A previously interrupted Marmousi execution retained its model, optimizer state, and 30 already charged gradients; resumed work did not reset its scientific budget. These execution amendments are disclosed because reproducibility includes the derivative and batching path, not only the final velocity arrays.

Input identity, configuration hashes, exact central-model initialization, budget counters, final arrays, numerical status, and independent metric recomputation were checked across the archive. These engineering checks strengthen the comparison's internal consistency. They do not remove target exposure, supply missing statistical replication, or justify a causal mechanism that the experimental design did not isolate.

%% file: data/all_metrics_table.tex
\begin{adjustbox}{max width=\linewidth}\begin{tabular}{llrrrr}
\toprule
Target & Method & RMSE & $\Delta_{\rm bg}$ (\%) & Train & Held out \\
\midrule
Marmousi region & Grid, direct & 237.80 & -13.64 & 1.01 & 1.18 \\
 & Grid, continued & 232.47 & -15.57 & 1.01 & 1.05 \\
 & SIREN, direct & 756.38 & +174.70 & 1.27 & 1.65 \\
 & SIREN, continued & 710.28 & +157.96 & 1.19 & 4.11 \\
 & Ours, direct & 223.23 & -18.93 & 1.01 & 1.14 \\
 & Ours, continued & 218.69 & -20.58 & 1.00 & 1.09 \\
\midrule
CurveVel-a01 & Grid, direct & 166.34 & -10.42 & 1.03 & 1.23 \\
 & Grid, continued & 100.26 & -46.01 & 1.01 & 1.05 \\
 & SIREN, direct & 615.62 & +231.52 & 1.15 & 2.85 \\
 & SIREN, continued & 826.15 & +344.89 & 1.77 & 5.95 \\
 & Ours, direct & 158.56 & -14.62 & 1.02 & 1.25 \\
 & Ours, continued & 107.97 & -41.86 & 1.01 & 1.17 \\
\midrule
CurveFault-a01 & Grid, direct & 537.67 & +22.10 & 1.34 & 2.12 \\
 & Grid, continued & 412.86 & -6.24 & 1.22 & 1.73 \\
 & SIREN, direct & 614.51 & +39.55 & 1.52 & 2.72 \\
 & SIREN, continued & 727.69 & +65.26 & 1.32 & 13.86 \\
 & Ours, direct & 511.23 & +16.10 & 1.18 & 11.14 \\
 & Ours, continued & 341.86 & -22.36 & 1.11 & 2.01 \\
\bottomrule
\end{tabular}\end{adjustbox}